\documentclass[aps,pra,amsmath,amssymb,twocolumn]{revtex4}
\usepackage{bbm}
\usepackage{graphicx}
\usepackage{dcolumn}
\usepackage{bm}
\usepackage{amsmath}
\usepackage{amssymb,amsthm}
\usepackage{color}
\usepackage{young}
\usepackage[sans]{dsfont}
 \usepackage{multirow}

\newcommand{\ket}[1]{|#1\rangle}
\newcommand{\braket}[2]{\langle{#1}|{#2}\rangle}

\newcommand{\bra}[1]{\langle#1|}

\def\eq{\begin{eqnarray}}
\def\en{\end{eqnarray}}

\def\beq{\begin{eqnarray}}
\def\een{\end{eqnarray}}
\def\bfig{\begin{figure}}
\def\efig{\end{figure}}

 \usepackage[normalem]{ulem}

\begin{document}

\title{Extending bosons and fermions beyond pairwise exchange symmetry}
\author{Malte C. Tichy and Klaus M\o{}lmer}
\address{Department of Physics and Astronomy, University of Aarhus, DK--8000 Aarhus C, Denmark}

\begin{abstract}
We study quantum many-body states of \emph{immanons}, hypothetical  particles that obey an exchange symmetry defined for more than two participating particles. Immanons thereby generalize bosons and fermions, which are defined  by their behavior under pairwise  symmetric and anti-symmetric exchange processes. The scalar product of two many-body states with  fermionic,  bosonic or generalized exchange symmetry becomes the determinant, permanent or immanant of the matrix containing all mutual scalar products of the occupied single-particle states.  As a measurable consequence,  immanons are shown to obey a partial Pauli principle that forbids the multiple occupation of single-particle states above a certain threshold. The tendency to favor or oppose multiple occupation of single-particle modes, i.e.~the degree of bunching, is the determinant, permanent or immanant of a hermitian positive semi-definite matrix. We exploit this identity to devise a Gedankenexperiment that corroborates the permanental dominance conjecture.
\end{abstract}
\date{\today}

\maketitle

\section{Introduction}
 The permutation symmetry of many-body wavefunctions, which is rooted in the fundamental indistinguishability of particles of the same species \cite{PauliNobel,Girardeau:1965ys}, is probably the most powerful symmetry principle in Nature. The  bosonic (fermionic) character of a wavefunction of many indistinguishable particles is reflected in the invariance (up to a sign) under the pairwise exchange of any two particle labels.
Enforcing this pairwise symmetry for the entire many-particle state and fixing the occupation numbers of all available single-particle modes results in a uniquely defined fully symmetric bosonic or fully anti-symmetric fermionic many-body state. 
The fermionic Pauli principle and the bosonic tendency to bunch constitute some of the empirically best verified physical principles \cite{Javorsek:2000bh,Collaboration:2010ly}, and they  explain a plethora of phenomena ranging from the statistics of thermal photons to the stability of matter.  Besides the abundance of natural  phenomena owing their existence to exchange symmetry, the degree of control over many individual particles allows a direct manipulation of synthetic exchange symmetry in the laboratory \cite{PhysRevLett.108.010502}, promising even to deliberately switch bosons into fermions \cite{Switchable2016} and  to simulate anyons \cite{Keilmann:2011aa}. Hence, artificial exchange symmetries  may not only deepen our understanding of the origin of exchange symmetry in nature, but also hold the potential for experiments and applications. 

In the mathematical treatment of anti-symmetric many-fermion wavefunctions, determinants appear ubiquitously \cite{Slater1929}, while
permanents \cite{permanents1965,Minc:1984uq,cheonUpdate}  -- the signless variant of the determinant --
 constitute the main workhorse when dealing with exchange-symmetric bosons \cite{Scheel:2004uq}.
From the perspective of character theory, however, the bosonic permanent and fermionic determinant are merely two special cases of a general complex-valued function on matrices, the immanant \cite{marcusMinc}:
  \eq
  \text{imm}(M) = \sum_{\sigma \in S_N} \prod_{j=1}^N  \chi_{\lambda}(\sigma) M_{j, \sigma_j} , \label{firstimmanant}
  \en
  where $\chi_{\lambda}(\sigma) $ is the \emph{character} labelled by the integer partition $\lambda$. The character is a particular function on the permutations $\sigma \in S_N$, which will be introduced formally below. For the trivial character $\chi(\sigma)=1$, we obtain the bosonic permanent, for the alternating $\chi(\sigma)=\text{sgn}(\sigma)$, the fermionic determinant. The relationship between the determinant, permanent and immanant of a matrix is an active area of mathematical research \cite{soules1994approach,Goulden1992,buergisser,cheonUpdate}.

Formally, definition (\ref{firstimmanant}) seems to naturally provide us with a  bridge between bosons and fermions, but the question arises to which extent the immanant can be given any physical interpretation beyond the bosonic and fermionic cases. So far, immanants have  only had rare appearances in physics; one application consists in the decomposition of the wavefunction of partially distinguishable bosons into components with different degrees of exchange symmetry   \cite{de-Guise:2014yf,Tan:2013ix,Tillmann:2014ye}. In this context,  states with immanonic symmetry have thus far always  appeared in   conjunction with states respecting other symmetries, and not for their own sake.

Here, we study many-body states that obey non-pairwise exchange symmetries. For these states, physical observables naturally take the form of immanants, promoting the formal construct Eq.~(\ref{firstimmanant}) to a  physical quantity. We show that immanants obey a partial Pauli principle which allows certain occupation multiplicities, while forbidding others. Eventually, we  generalize results of Ref.~\cite{TichyMultidimperm} and propose a (Gedanken-)experiment in which the immanant of a hermitian positive semidefinite matrix becomes the degree of bunching of a system of $n$ non-interacting interfering partially distinguishable particles.

\begin{figure}[th]
\includegraphics[width=\linewidth]{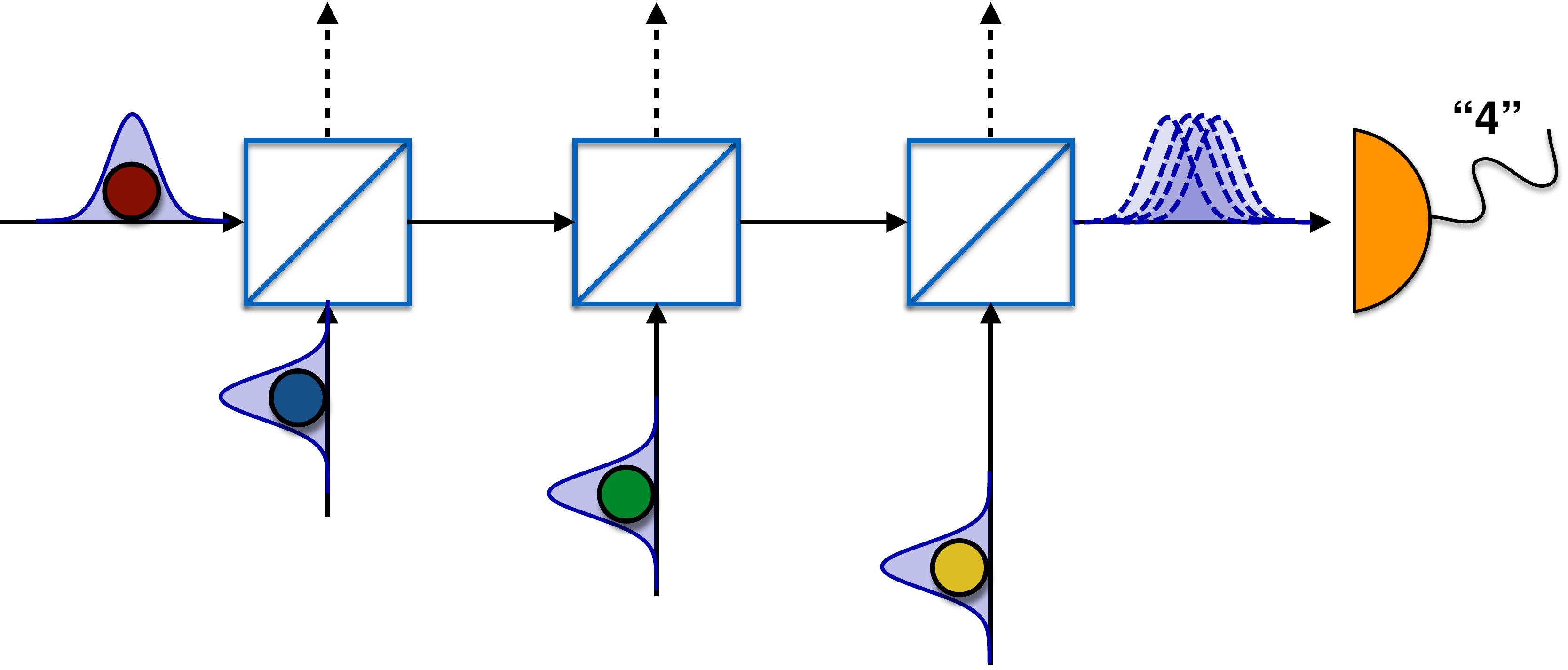}
\caption{Bunching experiment. Particles fall onto a cascaded arrangement of beam splitters with certain reflectivities, such that the classical combinatorial expectation for the probability to find all particles in the detector is the product of the individual probabilities,  e.g.~$P_{\text{D}}(\text{4-fold coincidence})=P_{1} P_{2} P_{3} P_{4}$. For indistinguishable  quantum particles, the resulting probability $P_{\text{quantum}}$ is the classical probability $P_{\text{D}}$ multiplied by a bunching factor. As we show in this article, the latter is, in general, the immanant of the distinguishability matrix $\mathcal{S}$ which describes the mutual overlap of the single-particle wavefunctions occupied by the particles.} \label{sketchBunchingExperiment}
\end{figure}

The probability to find all $n$ particles in the same single-particle state after some interaction-free evolution (see Fig.~\ref{sketchBunchingExperiment}) is increased by a factor $n!$ for identical bosons \cite{PhysRevLett.111.130503} with respect to the combinatorial expectation, while the respective probability for identical fermions vanishes -- this is one  of several possible generalisation \cite{Tichy:2010ZT,Tichy:2012NJP,Laloe:2010uq,Lim:2005qt,Mhrlein2015,PhysRevA.91.013811})  of the two-particle bosonic \cite{Hong:1987mz} or fermionic \cite{Bocquillon01032013} Hong-Ou-Mandel effect. In practice, the factor $n!$ for bosons and zero for fermions is  modified when the particles are not fully indistinguishable \cite{Niu:2009pr}. Such partial indistinguishability can be treated in different ways  \cite{tichyTutorial,ShchesnovichPartial2014,Shchesnovich2015,PhysRevLett.114.243601,Tillmann:2014ye,Tan:2013ix,de-Guise:2014yf}; here, we generalize the approach of Ref.~\cite{TichyMultidimperm}, in which the bunching factor for bosons becomes the permanent of the \emph{distinguishability matrix} that contains the $n^2$ scalar products of the single-particle wavefunctions.

Anticipating our results below -- and maybe not surprisingly --, the degree of bunching for fermions becomes the determinant, for immanons the immanant of the distinguishability matrix. 
 In this context, the  physically plausible statement that bosons are the particle species that favor multiply occupied states most, while fermions strictly prohibit such multiple occupation, becomes equivalent to the widely discussed, yet unproven \emph{permanental dominance conjecture} \cite{Lieb1966,cheonUpdate}  for positive-definite hermitian matrices. Together with its already proven counterpart for determinants, Schur's inequality \cite{Schurinequality}, the permanental dominance conjecture then implies that the degree of bunching of any state of any symmetry can neither surpass bosons nor underbid fermions.

We introduce many-body states with generalized exchange symmetry in Section \ref{immanons}, for which we formulate and prove the partial Pauli principle in Section \ref{partialpauli}. Our main result, the bunching probability for a system of $n$ immanons, is derived in Section \ref{collectiveinterference} and related to the permanental dominance conjecture.  In our conclusions in Section \ref{conclusions}, we  identify other aspects of immanonic states deserving further exploration.

\section{Many-body states with generalized exchange symmetry} \label{immanons}
The $d^n$-dimensional many-particle Hilbert-space which accommodates quantum states of $n$ particles  in $d$ single-particle modes can be decomposed in subspaces that each obey a particular type of exchange symmetry.  The most familiar  exchange symmetries are the bosonic permutation symmetry (leaving the fully symmetric subspace invariant) and the fermionic alternating symmetry (characterizing the fully anti-symmetric subspace). Before proceeding to more general types of exchange symmetry, we establish some useful group-theoretic notions.

\subsection{Permutation characters}
We will extensively deal with the \emph{character} $\chi_{\lambda}(\sigma)$ of a permutation $\sigma \in S_N$  \cite{Bernstein:2009uq,James1993,MultilinearAlgebra,Collins}.
A character is a function on the permutations that belongs to an irreducible representation $\lambda$ of the symmetric group $S_n$,  defined by an integer partition $\lambda \equiv ( \lambda_1, \dots , \lambda_L)$ of $n$, i.e.~$\lambda_{1} \ge \lambda_{2} \ge \dots \lambda_L \ge 0$, where $L$ is the \emph{length} of $\lambda$ and $n =\sum_{j=1}^L \lambda_j $ is its \emph{size}. The character of the trivial identity permutation  $\chi_\lambda(e)$ (with $e(k)=k$) coincides with the dimensionality of the vector space of the group representation defined by $\lambda$ \cite{James1993,Collins}.  For non-trivial representations, we have
\eq
\chi_\lambda(e) = \frac{n!}{\prod_{j=1}^n h_j!} ,
\en
where the $h_j$ are the hook-lengths of the associated Young diagrams that illustrate the respective integer partition \cite{CombinatoryAnalysisMacmahon}. 

 Character functions are class functions, i.e.~two conjugated permutations $\sigma, \rho = \tau \sigma \tau^{-1}$ possess the same character, $\chi_\lambda(\sigma)=\chi_\lambda(\rho)$. Two permutations are mutually conjugated if they possess the same cycle structure \cite{Collins,James1993}. For example,
\eq (1,2)(3) \hat = (1,3)(2) \hat \neq (1,2,3) \hat \neq (1)(2)(3) \hat \neq  (1,2)(3) . \en
Since each cycle structure is associated to exactly one integer partition  (the lengths of the cycles add up to $n$), the number of different cycles coincides with the number of integer partitions $P_n$. Consequently, the total number of inequivalent character values $\chi_\lambda(\sigma)$ for a given $n$  is $ P_n^2$. These values can be listed in character tables \cite{Collins,James1993}, two of which are listed in Table \ref{charactertables}.
\begin{table}
\begin{tabular}{r|r|r}
 $N=2$                    &  $\sigma = (1,2)$ & $\sigma=(2,1) $  \\ \hline
$\lambda = (1,1) $& 1 & -1 \\
$\lambda = (2) $& 1 & 1
\end{tabular}
\begin{tabular}{r|r|r|r}
             $N=3$        &  $\sigma = (1,2,3)$ & $\sigma=(2,1,3) $ & $\sigma = (2,3,1)$  \\ \hline
$\lambda = (1,1,1) $& 1 & -1 & 1 \\
$\lambda = (2,1) $& 2 &0 & -1  \\
$\lambda = (3) $& 1 & 1 & 1
\end{tabular}
\caption{Character tables for $N=2$ and $N=3$. The first and last rows correspond to fermionic and bosonic symmetry, respectively. } \label{charactertables}
\end{table}
General recipes for the enumeration of characters can be found in Refs.~\cite{Collins,FultonHarris,James1993,EnumerativeCombi}; in this work, we used the package provided by \cite{Miszczak}.

\subsection{Exchange symmetry}
In order to study the behavior of wavefunctions that live in a subspace of the many-body Hilbert-space obeying a certain exchange symmetry, we introduce the permutation operator $\hat Q_\sigma$,
\eq
\hat Q_\sigma \ket{\psi_1, \dots , \psi_n} = \ket{ \psi_{\sigma_1}, \dots, \psi_{\sigma_n} }  , \label{Qop}
\en
which permutes the labels of the particles.  Using this operator, we can define the projector onto the subspace subjugated to the exchange symmetry imposed by the character $\chi_\lambda$ \cite{pierce1970orthogonal,merris1976structure},   \eq
\mathcal{\hat P}_{\lambda} = \frac{\chi_\lambda(e)}{n!} \sum_{\sigma \in S_n} \chi_\lambda(\sigma) \hat Q_\sigma . \label{immanizer}
\en
Using Schur's orthogonality relations \cite{Collins,FultonHarris,James1993},
\eq
\sum_{\sigma \in S_n}  \overline{ \chi_{\eta}(\sigma) } \chi_{\lambda} (\sigma \tau ) = \frac{\delta_{\lambda,\eta} n! }{ \chi_{\lambda}(e) }  \chi_{\lambda} (\tau) , \label{schurortho}
\en
we find that the $\mathcal{\hat P}_\lambda$ indeed project onto orthogonal subspaces \cite{pierce1970orthogonal}
\eq
\mathcal{\hat P}_{\lambda} \mathcal{\hat P}_{\eta} & = & \mathcal{\hat P}_{\lambda} \delta_{\lambda,\eta} , \label{orthoproj}
\en
and that they span the full $d^n$-dimensional Hilbert space \cite{pierce1970orthogonal}: 
\eq  
\sum_{\lambda} \mathcal{\hat P}_\lambda   & = &
\sum_{\lambda} \sum_{\sigma \in S_n}  \chi_\lambda(e) \chi_\lambda(\sigma) \frac{1}{n!} \hat Q_\sigma  \\
&=& \sum_{\sigma \in S_n} \frac{1}{n!}  \hat Q_\sigma  \sum_{\lambda}  \chi_\lambda(e) \chi_\lambda(\sigma)  \\
&=& 
 \hat Q_{e} \equiv \mathbbm{1},
\en
where we used $\chi_\lambda(e) =\bar \chi_\lambda(e) $ and the character orthogonality relation for columns,
\eq
\sum_{\lambda}  \bar \chi_\lambda(e) \chi_\lambda(\sigma) =n!  \delta_{e,\sigma}
.\en
The orthogonality of symmetrizers pertaining to different partitions motivates us to consider each partition as one \emph{species}, with $\lambda=(n)$ denoting bosons and $\lambda=(1,1,\dots,1)$ denoting fermions.
Bosons implement the trivial constant character $\forall \sigma: \chi_{\lambda=(n)}(\sigma)=1$, the corresponding symmetrizer $\mathcal{\hat P}_{(n)}$ projects onto the ${d+ n-1 \choose n}$-dimensional fully symmetric subspace of the many-body Hilbert-space.  Since the symmetrizer $\mathcal{\hat P}_{(n)}$ is unaffected by any additional permutation, $\forall \sigma: \mathcal{\hat P}_{(n)} \hat Q_\sigma = \hat Q_\sigma \mathcal{\hat P}_{(n)} =\mathcal{\hat P}_{(n)} $, bosonic states are eigenstates  with the eigenvalue unity of all permutation operators $\hat Q_\sigma$.

Fermions transform by a one-dimensional representation given by the alternating character $\chi_{\lambda=(1,1, \dots, 1)}(\sigma)=\text{sgn}(\sigma)$, the operator  $\mathcal{\hat P}_{(1,1, \dots , 1)}$ is the anti-symmetrizer. The dimension of the anti-symmetric subspace is ${d \choose n}$, i.e.~there is no anti-symmetric state of $n$ particles that occupy  $d<n$ single-particle states. Due to $\mathcal{\hat P}_{(1, \dots, 1)} \hat Q_\sigma  = \text{sgn}(\sigma) \mathcal{\hat P}_{(1, \dots, 1)}$, all fermionic states are eigenstates of all possible permutation operators, just like bosons, but their eigenvalues coincide with the signature (+1 or -1) of the respective  permutation.

Pairwise exchange symmetry of two particles is restricted to symmetry or anti-symmetry and leaves no room beyond bosons and fermions. For  $n>2$, however, there are further irreducible, higher-dimensional representations of the symmetric group $S_n$ that come with their associated characters, one for each integer partition $\lambda$.

Even though no elementary physical particle is known that naturally implements a symmetry related to a character that is neither trivial nor alternating, we can always think of artificially preparing $n$  particles in  states obeying such symmetries. This may, e.g., be achieved by using an auxiliary internal degree of freedom to symmetrize the wave-function appropriately \cite{Matthews:2013aa,Shchesnovichfermions}. We will refer to particles that fulfil a generalized exchange symmetry, i.e. that live in the eigenspace of $\mathcal{P}_{(\lambda)}$ as \emph{immanons}.

\subsection{Immanons}
A many-immanon-state, i.e.~an eigenstate of $\mathcal{\hat P}_\lambda$ [Eq.~(\ref{immanizer})], 
 will remain in the subspace fulfilling the exchange symmetry defined by $\lambda$ if the time-evolution is induced by a permutation-symmetric Hamiltonian
\eq
\forall \sigma: [ \hat H, \hat Q_\sigma ] = 0 ,
\en
since such  Hamiltonians then also fulfil
\eq
 [ \hat H, \mathcal{\hat P}_\lambda ] = 0 .
\en
In other words, if the particles are treated in an indistinguishable manner, they retain the initially obeyed exchange symmetry. This generalizes the well-known invariance of bosonic and fermionic exchange symmetry under any time-evolution that respects the exchange symmetry: Due to their physical indistinguishability, bosons remain bosons and fermions remain fermions.

We will deal with symmetrized states,
\eq
\ket{\Psi_\lambda }&=&
 \frac{\sqrt{n!}}{\chi_\lambda(e)} \mathcal{\hat P}_{\lambda} \ket{\psi_1, \dots, \psi_n}, \label{firstimmanonstate} \\
&=& \frac{1}{\sqrt{n!}} \sum_{\sigma \in S_n} \chi_{\lambda}(\sigma) \left[ \otimes_{j=1}^n \ket{\psi_{\sigma_j}}_j \right]
  \label{genst}
\en
where the normalization factor is chosen such that the symmetrizer applied onto a separable  state in which each particle initially occupies a different state,
\eq
\mathcal{\hat P_\lambda} \ket{\psi_1,\psi_2, \dots, \psi_n},
\en
is normalized to unity thanks to
\eq
\sum_{\sigma \in S_n} | \chi_{\lambda}(\sigma) |^2 = n! ,
\en
for any character $\chi_{\lambda}$.

\subsection{Dependence of many-immanon-states on the seed}
It is natural to ask for a Fock-state-like representation of immanonic states. For bosons and fermions, the occupation of each mode is sufficient to fully characterize the state, since
\eq
\mathcal{\hat P}_{\lambda=(n) }  & = & \mathcal{\hat P}_{\lambda=(n) }  \hat Q_\sigma  \\
\mathcal{\hat P}_{\lambda=(1,\dots, 1) }  & = &  \text{sgn}(\sigma)  \mathcal{\hat P}_{\lambda=(1,\dots, 1) }  \hat Q_\sigma  ,
\en
i.e.~it is irrelevant whether we symmetrize a state $\ket{\Psi}$ or first apply any permutation $\sigma$ onto that state. For example, we can use both $\ket{a,b}$ and $\ket{b,a}$ as ``seed'' for the two-particle fermionic or bosonic Fock-states. Therefore, we can safely state the occupation of each mode in an arbitary order and use Fock-states for this purpose.

Many-immanon states, however,  depend on the initial ordering of the single-particle states, since for $\lambda \neq (1,\dots, 1) , (n)$,
\eq
\nexists f(\sigma) \in \mathbb{R}: \mathcal{\hat P}_{\lambda}  = f(\sigma) \mathcal{\hat P}_{\lambda}  \hat Q_\sigma  , \label{notinvariantP}
\en
for non-trivial permutations $\sigma$. This rules out a simple representation as Fock-states, and it remains to be studied how many-immanon-states can be represented in an  efficient way.

\subsection{Broken pairwise symmetry}
A state that is (anti)-symmetric upon the exchange of two particles,
\eq
\hat Q_{\sigma = (k,j)} \ket{\Psi}  = \pm \ket{\Psi},
\en
cannot be the eigenstate of an immanonic symmetrizer with $\lambda \neq (1, \dots, 1), (n)$, due to (\ref{notinvariantP}). That is to say, despite being the eigenstate of a weighted sum of permutation operators, non-trivial and non-alternating many-immanon-states do not remain invariant upon the exchange of any two particles.

\subsection{Immanants as many-body scalar product}
We have introduced classes of many-body wavefunctions that obey a certain exchange symmetry; this will now allow us to obtain a physical interpretation of the the immanant of a matrix in close analogy to the permanent and determinant. For that purpose, we consider the scalar product of two symmetrized $n$-body states,
\eq
\ket{\Psi_{\lambda}} &=& 
 \frac{\sqrt{n!}}{\chi_\lambda(e)} \mathcal{\hat P}_{\lambda} \ket{\psi_1, \dots, \psi_n},   \\
\ket{\Phi_{\lambda}} &=&  \frac{\sqrt{n!}}{\chi_\lambda(e)} \mathcal{\hat P}_{\lambda} \ket{\phi_1, \dots, \phi_n} .
\en
Due to the orthogonality of symmetrizers, Eq.~(\ref{orthoproj}), immanonic states related to different representations, i.e.~different integer partitions $\lambda, \eta$, are orthogonal,
\eq
\lambda  \neq \eta \Rightarrow \braket{\Phi_{\lambda}}{\Psi_\eta} =0 .
\en
The scalar product of two $n$-immanon-states of the same species, i.e.~the same partition $\lambda$, is
\eq
\braket{\Phi_{{\lambda}}} {\Psi_{{\lambda}}} & = &\frac 1 {n!} \sum_{\sigma, \rho \in S_n} \overline{  \chi_{\lambda}(\sigma)}  \chi_{\lambda}(\rho) \prod_{j=1}^n \braket{\phi_{\sigma_j} }{\psi_{\rho_j}}  \label{scalprgenst} \\
&=& \frac 1 {n!} \sum_{\tau \in S_n} \left( \sum_{\sigma \in S_n}  \overline{ \chi_{\lambda}(\sigma) } \chi_{\lambda}(\sigma \tau ) \right) \prod_{j=1}^n \braket{\phi_{j} }{\psi_{ \tau_j}} \nonumber  \\
& = &   \frac{1 }{\chi_{\lambda}(e)} ~ \text{imm}_{\lambda}(M) , \label{immaeq}
\en
where we used the Schur orthogonality relations (\ref{schurortho}), 
defined  the matrix
\eq
M_{j,k} = \braket{\phi_j}{\psi_k} , \label{Metaeps}
\en
and eventually recovered the immanant (\ref{firstimmanant}).

In other words,   the scalar product of  two many-immanon-states of the same species is the immanant of the matrix that contains all the single-particle scalar products $\braket{\phi_j}{\psi_k}$. We can revert this argument: Via the singular-value-decomposition of an arbitrary matrix $M$, $M= U^{-1} D V$ with $U,V$ unitary and $D$ diagonal, single-particle quantum states $\ket{\phi_j}$ and $\ket{\psi_j}$ can be chosen such that Eq.~(\ref{Metaeps}) is fulfilled, i.e.~any immanant can be written as the many-body scalar product of two many-immanon-states.

Both the trivial and the alternating representations are one-dimensional, $\chi_{(n)}(e)=\chi_{(1,1, \dots ,1)}(e)=1$, and we recover the well-known relations
\eq
\braket{\Phi_{(n)}} {\Psi_{(n)}} &= &  \text{perm}(M) , \\
\braket{\Phi_{(1,\dots, 1)}} {\Psi_{(1,\dots, 1)}} &=&  \text{det}(M) ,
\en
where $\text{perm}$ denotes the permanent \cite{Minc:1984uq} for bosons ($\chi_{(n)}(\sigma)=1$) and $\text{det}$ the determinant for fermions ($\chi_{(1,\dots, 1)}(\sigma)=\text{sgn}(\sigma)$).

\section{Collective many-immanon interference} \label{collectiveinterference}
\subsection{The partial Pauli principle for immanons} \label{partialpauli}
Before proceeding to the dynamics of interfering immanons, we need to establish which many-immanon states can be populated at all. We are used to encounter either the full suppression of multiply occupied states or no suppression whatsoever. The Pauli principle entails  a  vigorous boundary condition on the set of accessible quantum states for fermions: No two identical fermions can populate the same single-particle state. This physical statement is, both, origin and consequence of the exchange anti-symmetry of the many-fermion wavefunction. On the other hand, postulating that all particles \emph{can} occupy the same state is an equally strong postulate. Again, it is equivalent to a fully symmetric, bosonic wavefunction. Both principles can already be formulated for two particles and indeed immediately imply pairwise exchange symmetry. For two particles, there are no other possible occupation rules: Double occupation is either forbidden (for two fermions) or allowed (for two bosons). Many-body states of fermions and bosons thus directly inherit the fermionic Pauli principle and the bosonic possibility to multiply occupy single-particle states.

For systems of more than two particles, however, we may imagine a weaker form of the Pauli principle, an intermediate form of occupation rules, which, for example, allows the double- but not the triple- or higher population of a single-particle state. Anticipating this Section's results, immanons precisely implement such a \emph{weak} or \emph{partial} Pauli principle.

Applying the symmetrizer $\mathcal{\hat P}_{\lambda} $ related to a partition $\lambda$ to a state with multiple occupations, we find a hierarchy of allowed multiplicities for immanons, that is, a ladder between the strong Pauli principle for fermions and the absence of any  rules restricting multiple occupations for bosons. To be more precise, let us express the population multiplicities of $N$ particles as integer partitions. Two population multiplicities $\lambda$ and $\eta$ can be ordered partially via the majorization criterion:
\eq
\forall m: \sum_{j=1}^m  \lambda_j \le \sum_{j=1}^m  \eta_j  \Leftrightarrow \lambda \precsim \eta   .
 \en
 The induced order is only partial: We have, e.g., $(1,1,1,1) \precsim (2,1,1) \precsim (2,2) \precsim (3,1) \precsim (4)$, but neither $(2,2,2) \precsim (3,1,1,1)$ nor $(2,2,2) \succsim (3,1,1,1)$.

The partial order of integer partitions dictates the generalized Pauli principle for immanons: Immanons that obey the exchange symmetry induced by the partition $\lambda$ can occupy all states with population multiplicities $\eta$  fulfilling  $\eta \precsim \lambda$.

Mathematically speaking, this statement is equivalent to
\eq
\mathcal{\hat  P}_{\lambda} \ket{ \Psi }_\eta \neq 0 \Rightarrow \eta \precsim \lambda,  \label{formulatepauli}
\en
for all states of the form $\ket{\Psi}_\eta= \otimes_{j=1}^n \ket{j}^{\otimes k_j} $ with $\eta=(k_1, k_2, \dots , k_n)$, i.e. states with mode occupation numbers $k_j$ corresponding to the integer partition $\eta$.

For bosons, the generalized Pauli principle implies no restriction at all, since $\eta \precsim (n)$ for all partitions $\eta$, while it reduces to the usual strong Pauli principle for fermions, since $ (1,1,\dots ,1) \precsim \eta$ for all $\eta$.

The validity of the partial Pauli principle can be shown formally using methods from character theory worked out in Ref.~\cite{Brylinksi}. In order to obtain an intuitive picture, let us  consider the unique many-body state for which all particles occupy the same single-particle state $\ket{\phi}$, \eq \ket{\text{BEC}} = \otimes_{j=1}^n \ket{\phi} . \en
This state is the unique eigenstate of $\mathcal{\hat P}_{(n)}$. Then, due to the orthogonality of projectors (\ref{orthoproj}), the projection onto the subspace associated to any other character vanishes,
\eq
\mathcal{\hat P}_{\lambda \neq (n) } \ket{\text{BEC}} = 0 .
\en
That is, bosons are the only species for which all particles can occupy the same single-particle state. This argument follows the lines of the proof  of Lemma 5.1 in Ref.~\cite{Brylinksi}, and it can be extended to show that (\ref{formulatepauli}) holds in general.

\subsubsection{Non-locality of the partial Pauli principle}
The partial Pauli principle is a highly non-local phenomenon: Consider the four-particle immanonic state $\mathcal{\hat P}_{\lambda=(2,1,1)} \ket{a,a,c,d}$, where $\ket{a}\dots \ket{d}$ are mutually orthogonal single-particle states. Focusing on the particles in $\ket{c}$ and $\ket{d}$, these will behave as two fermions, since they cannot doubly occupy any single-particle state, e.g. $\mathcal{\hat P}_{\lambda=(2,1,1)} \ket{a,a,c,c}$ and $\mathcal{\hat P}_{\lambda=(2,1,1)} \ket{a,a,d,d}$  both vanish identically due to the partial Pauli principle. Hence, a coupling between $\ket{c}$ and $\ket{d}$, e.g.~by a Hamiltonian of the form $\hat H = - J (\ket{c} \bra{d} + \ket{d} \bra{c})$ with $J$ being the coupling strength, will never induce the multiple population of neither $\ket{c}$ nor $\ket{d}$.

When, however, the initial state is $\mathcal{\hat P}_{\lambda=(2,1,1)} \ket{a,b,c,d}$, the two particles in $\ket{c}$ and $\ket{d}$ can both occupy $\ket{c}$ or both occupy $\ket{d}$ after the time-evolution mediated by the coupling Hamiltonian - the partial Pauli principle does not forbid such dynamics.

\begin{figure}[th]
\includegraphics[width=\linewidth]{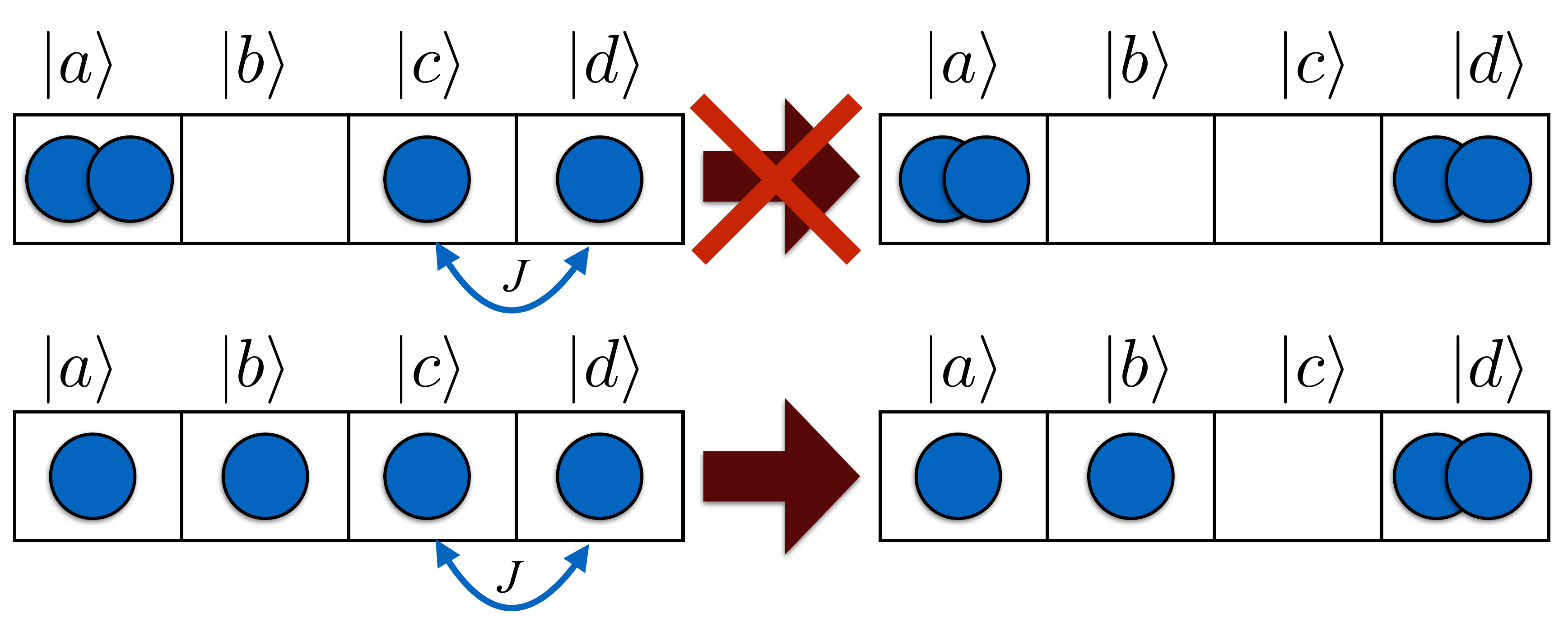}
\caption{
Non-locality of the partial Pauli principle. Upper panel: For the arrangement of particles $\mathcal{\hat P}_{\lambda=(2,1,1)} \ket{a,a,c,d}$, the particles initially occupying the states $\ket{c}$ and $\ket{d}$ cannot both end up in the same single-particle state, even in the presence of coupling, due to the partial Pauli principle: $(2,1,1) \precsim (2,2)$. In the lower panel, the double occupation of $\ket{d}$ is possible, since no other mode is initially doubly occupied.
 } \label{figweakpauli}
\end{figure}

That is, the occupation of the first two modes governs the dynamics of the particles occupying the other two modes, as sketched in Fig.~\ref{figweakpauli}, even though the modes $\ket{a}, \ket{b}$ might be spatially separated from the modes $\ket{c}, \ket{d}$. In other words, the statistical behavior of the particles in $\ket{c}$ and $\ket{d}$ can be steered by the multiple occupation of $\ket{a}, \ket{b}$. Given that non-interacting fermions have no non-trivial quantum-computational power \cite{Valiant}, while non-interacting bosons allow one to perform measurement-based quantum computation \cite{Knill:2001aa}, the question naturally arises to which extent immanons can be exploited for the design of quantum gates, possibly by exploiting the non-local steering property discussed above.

\subsection{Many-particle scattering}
\begin{figure}[th]
\includegraphics[width=\linewidth]{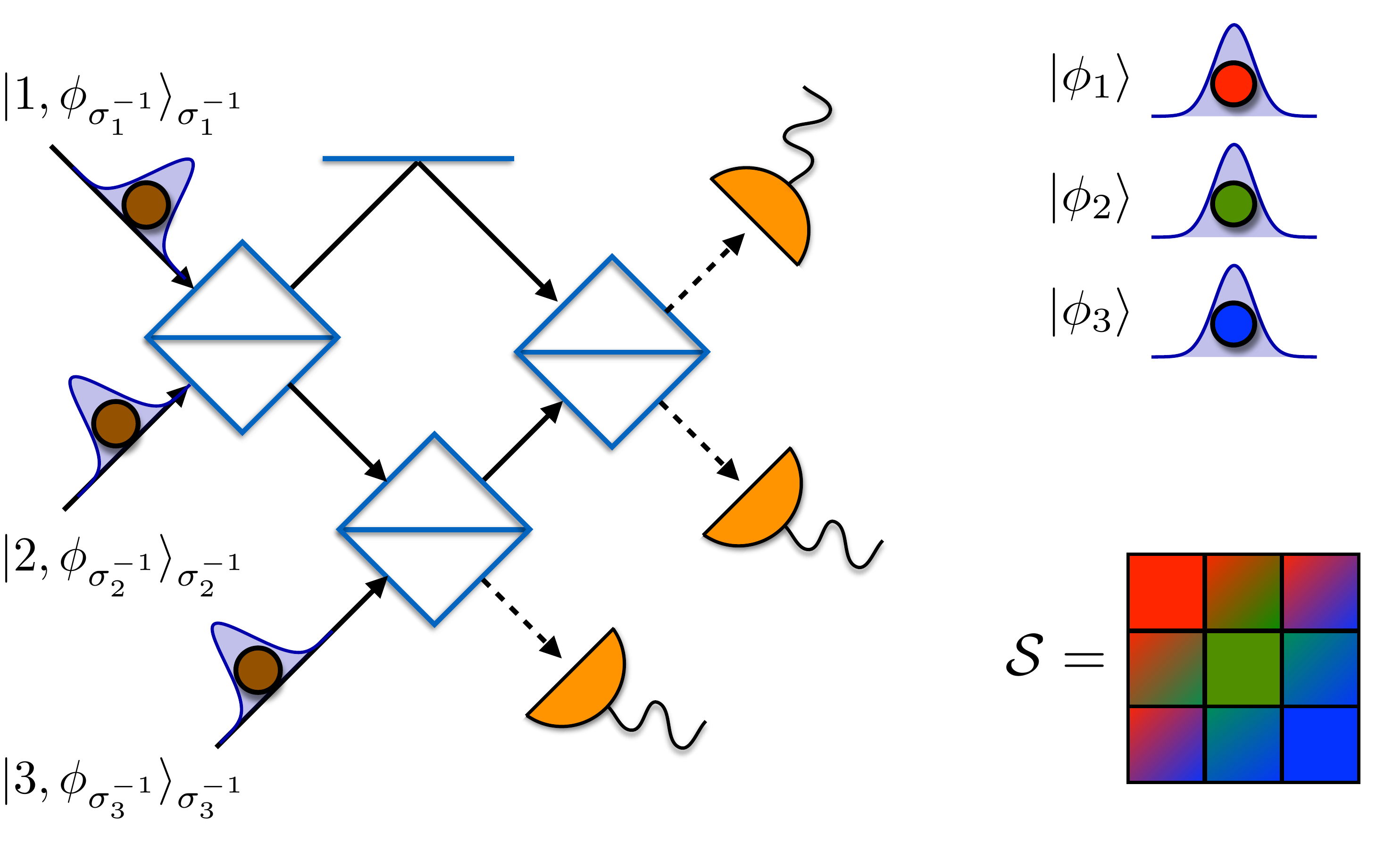}
\caption{
Multi-mode scattering setup for $n=3$ particles.  The particles carry the internal states $\ket{\phi_1}, \ket{\phi_2}, \ket{\phi_3}$, the initial state is a superposition of all permutations $\sigma$ of the three particles in the three modes, Eq.~(\ref{inistate}). The particles are detected by the three detectors after the beam splitters. Any unitary transformation $U$ can be realized by beam splitters (possibly with additional phase shifters) \cite{Reck:1994zp}. The probability to find the particles in the output modes depends on the internal states, as expressed by the distinguishability matrix $\mathcal{S}$, the symmetry of the state encoded in $\lambda$, and the scattering setup $U$.  
 } \label{multimode.pdf}
\end{figure}

So far, we have focused on rather static situations and described the formal properties of many-immanon states. With the help of a dynamical setup in which partially distinguishable immanons interfere, the  physical meaning of the immanant of unitary and of positive semi-definite matrices becomes apparent. For this purpose, we closely follow the scattering setup of Ref.~\cite{TichyMultidimperm} and generalize it to immanons. We consider
 $n$ immanons prepared in the state
\eq
\ket{\Psi_{\text{ini}}} = \frac{1}{\sqrt {n!}} \sum_{\sigma \in S_n}  \chi_\lambda(\sigma) \otimes_{j=1}^n \ket{ \sigma_j, \phi_{\sigma_j} }_j  , \label{inistate}
\en
where  $\ket{m, \phi_p }_q $ denotes the state of the $q$th particle prepared in the $m$th external mode and in the internal state $\ket{\phi_p}$. The physical situation is sketched in Fig.~\ref{multimode.pdf}. 

We assume $\braket{ m , \phi_p}{ n , \phi_q} = \delta_{m,n}  \braket{ \phi_p}{\phi_q}$, the state (\ref{inistate}) is therefore properly normalized to unity. We store the mutual distinguishability of the particles with respect to their internal degree of freedom in a positive-definite hermitian matrix, the distinguishability matrix \cite{TichyMultidimperm}
\eq
\mathcal{S}_{j,k} = \braket{\phi_{j} }{\phi_k} .
\en

The particles scatter off  a linear setup $M$ (which is not necessarily unitary \cite{tichy:2013PRA}), which induces the following evolution on the level of single-particle states:
\eq
\ket{j , \phi_j} \rightarrow \sum_{k=1}^n M_{j,k} \ket{k, \phi_j} , \label{singletimeevo}
\en
i.e.~the external degree of freedom $\ket j$ evolves into a superposition $\sum_k M_{j,k} \ket{k}$, while the internal degree of freedom $\ket{\phi_j}$ remains unaffected. This corresponds, for example, to the time-evolution of photons in a multi-mode array, where the polarization remains unaffected by the propagation through the network.

The time-evolution (\ref{singletimeevo}) can be inserted into the many-body state (\ref{inistate}), to yield
\eq
\ket{\Psi_{\text{evo}}} = \frac{1}{\sqrt {n!}} \sum_{\sigma \in S_n}  \chi_\lambda(\sigma) \otimes_{j=1}^n \sum_{k=1}^n M_{\sigma_j, k} \ket{k , \phi_{\sigma_j} }_j   \label{evolvedstate} ,
\en
i.e.~a superposition of all possibilities to distribute the particles among the modes.
We now focus on an event with exactly one particle in each external mode, i.e.~we project the state onto
\eq
\mathcal{\hat Q}_{\text{coin}} = \sum_{\sigma \in S_n} \sum_{x_1, \dots, x_n} \otimes_{j=1}^n \ket{ \sigma_j, x_j }  \bra{ \sigma_j, x_j } ,
\en
where the sum over $x_1, \dots, x_n$ takes into account all possible internal states. Identifying the relevant terms, we find
\eq
\ket{\Psi_{\text{proj}}} & = & \mathcal{\hat Q}_{\text{coin}} \ket{\Psi_{\text{evo}}} \\
&=&
 \frac{1}{\sqrt{n!}} \sum_{\sigma \in S_n} \chi_\lambda(\sigma) \sum_{\rho \in S_n} \otimes_{j=1}^n \left( M_{\sigma_j, \rho_j } \ket{\rho_j, \phi_{\sigma_j}}_j \right)  , \nonumber
\en
where we sum over all components of the state (\ref{evolvedstate}) with one particle per external mode. The probability $P_{\lambda}^{{\text{coin}}}$ to find one particle in each output mode is then the norm of this projected state,
\eq
P_{\lambda}^{\text{coin}} \hspace{-0.2cm}&=& \hspace{-0.1cm}
\braket{\Psi_{\text{proj}}}{\Psi_{\text{proj}}} \nonumber  \\
& =& \hspace{-0.35cm} \sum_{\bar \sigma, \sigma, \rho \in S_n} \frac{ \overline{\chi}_\lambda(\bar \sigma) \chi_\lambda(\sigma) }{n!}\prod_{j=1}^n M^*_{\bar \sigma_j, \rho_j}M_{\sigma_j, \rho_j} \mathcal{S}_{\bar \sigma_j, \sigma_j} \label{firstsum}
\\
& \hspace{-0.2cm}\stackrel{(
\ref{schurortho} )}{=}& \frac{1}{\chi_\lambda(e) } \sum_{\tau, \eta \in S_n} \chi_\lambda(\eta) \prod_{j=1}^n M^*_{j, \tau_j} M_{\eta_j, \tau_j} \mathcal{S}_{j,\eta_j}  \label{secondsum}
\en
which generalizes Eq.~(19) of Ref.~\cite{TichyMultidimperm} from bosons to general states with  exchange symmetry: Instead of dealing with a multi-dimensional permanent, we now deal with a multi-dimensional immanant. We can adapt the physical interpretation of \cite{TichyMultidimperm} as double-sided Feynman diagrams: A particle starts in mode $\rho_j$, ends in mode $j$, and travels ``back in time'' to mode $\tau_j$. This process carries the phase $\chi_\lambda(\tau^{-1} \rho)$ and it is weighted by $\mathcal{S}_{\tau_j, \rho_j}$.

The scattering into multiply occupied final states $\vec s=(s_1, s_2, \dots, s_n)$ is described by repeating the respective column of the scattering matrix $M$, and accounting for the multiple counting of identical events by a normalization factor $1/\prod_{j=1}^n s_j!$ \cite{TichyMultidimperm}, 
\eq
P_{\lambda}(\vec s) =  \frac{1}{\chi_\lambda(e) \prod_{j=1}^n s_j!} \times  \hspace{3cm} \nonumber \\  \hspace{2cm} \sum_{\tau, \eta \in S_n} \chi_\lambda(\eta) \prod_{j=1}^n M^*_{j, \tau_j} M_{\eta_j, \tau_j} \mathcal{S}_{j,\eta_j}   \label{multipleocc}
\en

\subsection{Fully indistinguishable particles}
For indistinguishable particles, $\mathcal{S}_{j,k}=1$, the sums over $\bar \sigma$ and $\sigma$ in Eq.~(\ref{firstsum}) become independent and we find for the probability to occupy the output particle arrangement $\vec s$, 
\eq
P_\lambda(\vec s) &=&  \frac{1}{n! \prod_{j=1}^n s_j!} \sum_{\rho \in S_n} \left| \sum_{\sigma \in S_n} \chi_\lambda(\sigma) \prod_{j=1}^n M_{\sigma_j, \rho_j} \right|^2   \label{immanantsample} \\
& =&\frac{1}{n! \prod_{j=1}^n s_j!} \sum_{\rho \in S_n}  \left| \text{imm}_\lambda(M_{\rho}) \right|^2 , \label{sumofimmanants}
\en
where $M_{\rho}$ denotes the matrix $M$ with columns permuted according to $\rho$. 
That is, the sum over absolute-squared immanants of the scattering matrix yields the event probability. Here, the immanants appear without the factor $1/\chi_\lambda(e)$. For bosons and fermions, the resulting permanent and determinant are invariant under the permutation $\rho$ of the columns of the matrix, and a single absolute-squared permanent or determinant emerges:
\eq
P_{\lambda=(n)}(\vec s) &=& \frac{1}{\prod_{j=1}^n s_j!}  \left| \text{perm}(M) \right|^2 , \\
P_{\lambda=(1,1,\dots, 1)}(\vec s) &=& \frac{1}{\prod_{j=1}^n s_j!}  \left| \text{det}(M) \right|^2 .
\en
During a scattering transition, the original and the partial Pauli principles appear dynamically: For multiply occupied final states $\vec s$ with multiplicity $\eta$, all immanants $\text{imm}_\lambda(M_{\rho})$ vanish identically if $\eta \precsim \lambda$ is not fulfilled. In particular, fermions never multiply occupy any mode, since the determinant of a matrix with any two or more identical columns vanishes.

\subsection{Fully distinguishable particles}
For distinguishable particles, $\mathcal{S}_{j,k}=\delta_{j,k}$, it is most convenient to use Eq.~(\ref{secondsum}), where only the summand with $\eta=e$ remains, and
\eq
P_{\text{D}}(\vec s) &=& \frac{1}{\prod_{j=1}^n s_j!}  \sum_{\sigma \in S_n} |M_{\sigma_j,j} |^2\\
& =& \frac{1}{\prod_{j=1}^n s_j!}  \text{perm}( |M|^2 ),  \label{classical}
\en
i.e.~the permanent of the matrix $|M|^2$, where the absolute-square is meant in an element-wise fashion. The exchange symmetry of the particles, inscribed in $\chi$, becomes irrelevant when the particles are prepared in fully distinguishable internal states. The distinguishability matrix $\mathcal{S}$ hence interpolates between the interference-dominated sum of immanants of the complex matrix $M$ Eq.~(\ref{sumofimmanants}), and the combinatorial permanent of the positive matrix $|M|^2$, Eq.~(\ref{classical}), which is independent of the particle species.

\subsection{Bunching events} \label{bunching}
For a bunching event with all particles in one (say, the first) output mode ($\vec s_{\text{bunch}}=(n, 0, \dots, 0)$), the scattering matrix has $n$ identical columns, and we use Eq.~(\ref{multipleocc}) to find
\eq
 P_{\lambda}(\vec s_{\text{bunch}}) = {P}_\text{D}(\vec s_{\text{bunch}})  \frac{\text{imm}_\lambda( \mathcal{S} ) }{\chi_\lambda(e)} ,  \label{immanantbunch}
\en
where
\eq
{P}_\text{D}(\vec s_{\text{bunch}}) = \prod_{j=1}^n |M_{j,1}|^2
\en
is the classical, combinatorial probability to find all particles in the first mode -- the product of the individual transition probabilities. Eq.~(\ref{immanantbunch}) thereby generalizes Eq.~(53) from Ref.~\cite{TichyMultidimperm} (formulated for bosons) to states with general exchange symmetry, including fermions.

The distinguishability matrix  $\mathcal{S}$ is constrained to $\mathcal{S}_{j,j}=1$. Since every positive semi-definite matrix $M$ can be written as
\eq
M = \mathcal{D} \mathcal{S} \mathcal{D}^*,
\en
where $\mathcal{D}$ is diagonal, and, thus,
\eq
\text{imm}_\lambda(M) = \text{imm}_\lambda(\mathcal{S}) \prod_{j=1}^n |\mathcal{D}_{j,j}|^2 ,
\en
 the immanant of every positive semi-definite matrix $M$ can be written (up to a factor) as the immanant of a distinguishability matrix $\mathcal{S}$.

Let us interpret the physical meaning of Eq.~(\ref{immanantbunch}): The probability that all particles end in the same mode depends, in the first place, on the combinatorial probability $P_{\text{D}}(\vec s_{\text{bunch}})$: Adapting the reflectivities of the beam-splitters in Fig.~\ref{sketchBunchingExperiment} will considerably affect the probability to find all particles in the detector. This classical expectation, however, is then modified by a factor that depends on the exchange symmetry \emph{and} on the distinguishability of the particles. The former is encoded in $\lambda$, the latter in $\mathcal{S}$.

Bosons tend to bunch, and the more the particles are indistinguishable, the stronger they will bunch. Fermions, on the other hand, tend to avoid multiple occupation of any single-particle state, the more strongly, the more indistinguishable they are. For bosons, bunching events are always enhanced with respect to the value for distinguishable particles, for fermions, they are always suppressed.

\subsection{Partially distinguishable particles}
\subsubsection{Bosonic and fermionic tendencies}
The tendencies to bunch and anti-bunch are directly reflected by Hadamard's inequality for the determinant and Marcus' inequality for the permanent of a positive semi-definite matrix, which for our $\mathcal{S}_{j,j}=1$ read
\eq
0 \le \det(\mathcal{S}) \le 1 \le \text{perm} (\mathcal{S}) \le n!
\en
The first and last inequalities are only saturated when $\forall j,k: \mathcal{S}_{j,k}=1$, i.e.~for fully indistinguishable particles. The generalization of Marcus' inequality by Lieb \cite{Lieb1966} concerns a positive semi-definite matrix with partition
\eq
\mathcal{S} = \left( \begin{array}{cc} A & B \\ B^\dagger & C \end{array} \right) ,
\en
we have
\eq
\text{perm}(\mathcal{S}) \ge \text{perm}(A) \text{perm}(C)  \ge \prod_{j=1}^n \mathcal{S}_{j,j} = 1,
\en
which physically means that two sets of partially distinguishable bosons that are all mutually distinguishable across the sets will bunch more when the sets are made more indistinguishable via the off-diagonal terms contained in $B$.

Fisher's inequality \cite{Fisher1907}
\eq
\text{det}(\mathcal{S}) \le \text{det}(A) \text{det}(C) \le \prod_{j} \mathcal{S}_{j,j} = 1 
\en
 gives us an analogous  interpretation for fermions: Any additional indistinguishability leads to a reduction in the multiple population of a single-particle state.

 The distinction between bosons and fermions could not be clearer: Their statistical behavior is perfectly opposite, with the combinatorial case as a benchmark lying in between.

\subsubsection{Permanental dominance and determinental subjugation}
Immanons can be compared to fermions by considering Schur's inequality \cite{Schurinequality} for positive semi-definite matrices from 1918: the minimum value (over $\lambda$) of the normalized immanant is attained by the determinant:
\eq
 \text{det}(\mathcal{S} ) \le \frac{ \text{imm}_\lambda(\mathcal{S} ) }{\chi_\lambda(e) }.
\en
That is, the alternating character minimizes the immanant for any positive-definite hermitian matrix. Fermions constitute the species that is most adverse to multiply populating a single-particle state. Fermions obey the Pauli principle without any exception, already the double occupation of any single-particle state is forbidden. Immanons, however, can tolerate the multiple population of single-particle states to a certain extent, hence the suppression of bunching events is not as strong as for fermions.

On the other hand, Lieb's permanental dominance conjecture \cite{Lieb1966} states that
 the maximum value (over $\lambda$) of the normalized immanant equals the permanent
\eq
\frac{\text{imm}_\lambda(\mathcal{S}) }{\chi_\lambda(e)}   \le \text{perm} (\mathcal{S})   , \label{permanentaldom}
\en
which physically implies that bosons constitute the species that favors multiply occupied states most. While many partial results are known for the permanental dominance conjecture
\cite{soules1994approach,pate1999tensor,cheonUpdate}, it remains  unproven in its generality.

 The exchange symmetry in the bosonic wavefunction pushes constructive interference for bunching events in the strongest foreseeable way \footnote{For elementary particles, composite bosons made of bosons may exhibit super-bosonic bunching behavior \cite{TichyTwoBosComp}.}. From a physics perspective, the permanental dominance conjecture is thereby a very plausible one: Bosonic bunching is the strongest bunching that can be attained with any exchange symmetry. Despite the abstract way of introducing the state (\ref{firstimmanonstate}), immanons can certainly be regarded as implementable in the laboratory. That is, a violation of the permanental dominance conjecture, i.e.~the existence of a positive-definite matrix $\mathcal{S}$ and a character $\lambda$ for which Eq.~(\ref{permanentaldom}) were violated, would imply a surprising physical situation in which partially distinguishable immanons bunch more strongly than partially distinguishable bosons.

\subsection{Distinguishability transition}
Given $n$ fully indistinguishable particles prepared in the state (\ref{inistate}), an $n$-fold population of a final state is only possible for bosons, and impossible for any other immanonic species. Depending on their character $\lambda$, immanons still favor multiply occupied states with a certain number of particles per mode. As a consequence, partially distinguishable immanons may feature enhanced bunching, while fully indistinguishable immanons are prohibited to all occupy the same state. This phenomenon can be appreciated with the help of the example \cite{TichyMultidimperm}
\eq
\mathcal{S}_{\text{transition}} = \left(\begin{array}{ccccc}
1 & x & x \dots & x \\
x &1 &  x \dots & x \\
\vdots & \vdots & \vdots & \ddots &\vdots \\
x &x&  x \dots & 1
\end{array}  \right) , \label{transitionmatrix}
\en
i.e.~each pair of two  single-particle mode functions possesses the same scalar product $x$.

\begin{figure}[th]
\includegraphics[width=\linewidth]{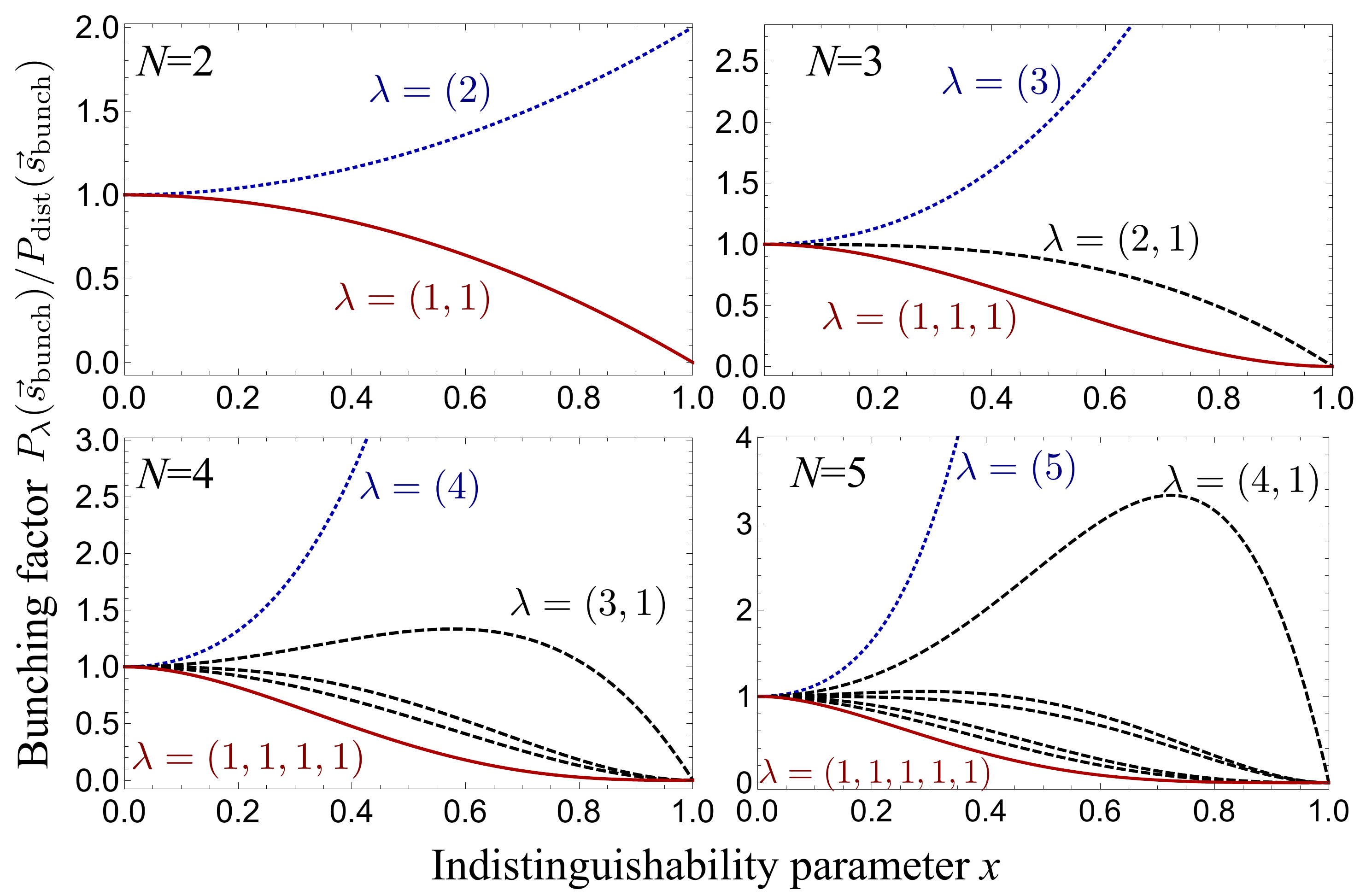}
\caption{
Bunching factor for partially distinguishable immanons described by the distinguishability matrix (\ref{transitionmatrix}). Blue dotted line: Bosons. Red line: Fermions. Dashed black lines: immanons corresponding to the respective symmetry defined by $\lambda$, the order of the lines reflects the natural majorization progression of the immanon-types: $(1,1,1,1)  \precsim (2,1,1)  \precsim (2,2)  \precsim (3,1)  \precsim ( 4) $ and $(1,1,1,1,1)  \precsim (2,1,1,1)  \precsim (2,2,1)  \precsim (3,1,1)  \precsim (3,2)  \precsim (4,1)  \precsim (5)$.
 } \label{FigureDistTransitionImmanons.pdf}
\end{figure}

We plot the different immanants of $\mathcal{S}_{\text{transition}}$ as a function of $x$ in Fig.~\ref{FigureDistTransitionImmanons.pdf}.
For $x=0$, the particles are fully distinguishable for any species, and there is no difference with respect to distinguishable particles. On the other hand, for $x=1$, the particles are perfectly indistinguishable; for such configuration of particles, only bosons are allowed to all populate the same state, hence all other immanants (including the determinant) vanish in this limit.

For bosons and fermions, the impact of indistinguishability is monotonic: The more indistinguishable many bosons are, the more they will tend to occupy the same output mode -- this monotonicity always persists for final states in which all particles occupy the same output mode but is broken for more general output arrangements \cite{Tichy2011,Ra:2013kx}. On the other hand, the more indistinguishable $n$ fermions are, the stronger they will avoid to all occupy the same mode.

Non-trivial immanons have a more ambivalent tendency: Since they may enhance multiply occupied output states -- as long as not all particles end in the  same mode -- multiple occupation can be enhanced for moderate values of partial distinguishability. Measuring all particles in one external output mode typically comes with finding some particles in the  same internal state and some in some other internal state. For certain $\lambda$ and intermediate values of $x$, we see neither  enhancement nor suppression with respect to distinguishable particles, similar to the interference patterns of many photons measured in final states other than the bunching arrangement \cite{Tichy2011,Ra:2013kx}.

\subsection{Defining many-particle coherence}
Distinguishability is a  measurable property since the bunching event probabilities in Eq.~(\ref{immanantbunch}), can, in principle, be retrieved in experiments. Distinguishability cannot, however,  be unambiguously defined accross particle species: Two matrices $\mathcal{S}$ and $\mathcal{T}$ can have similar determinants, yet  different permanents and immanants, or vice-versa. As a physical consequence, two many-body states with  similar degree of bosonic bunching can have a  different degree of fermionic anti-bunching. While pairwise distinguishability is clearly defined by the overlap of the two wavefunctions $\braket{\psi_1}{\psi_2}$, already the distinguishability of three particles becomes dependent of the particle species; in particular, it ceases to depend only on absolute values of the three  mutual scalar products \cite{threephotons2016}.

A mapping between the degree of bunching of bosons and the degree of anti-bunching of fermions is, hence, impossible, and so is a generic quantifier of many-particle distinguishability. The application of  recent approaches to the quantification of coherence \cite{Levi2014,Baumgratz2014} to the current situation seems, therefore, unfeasible.

\section{Conclusions and outlook} \label{conclusions}
Given the exotic properties that immanons fulfil,  theoretical physicists may be relieved that Nature seems to have chosen not to implement exchange symmetries related to such higher-dimensional representations of the symmetric group. Albeit a seemingly formal construct, immanonic states nevertheless constitute an interesting, yet challenging, field of study for a number of reasons: Immanons constitute the natural intermediate species between bosons and fermions, since they implement an intuitive form of a partial Pauli principle. The immanonic symmetrizers span the full many-body Hilbert-space and thereby saturate all possible states. Furthermore, they give a physical interpretation to a mathematical conjecture. The basic tools that ease our dealing with bosons and fermions remain to be generalized to the realm of many-immanon-states. Already the mere enumeration of basis states constitutes the most immediate desideratum.  We have also not yet enforced that immanons be \emph{indistinguishable} -- for bosons and fermions, enforcing pairwise exchange symmetry implies indistinguishability, which is not necessarily true for immanons. 
So far, we have restricted ourselves to a situation with constant particle number. Modelling the addition or subtraction of a particle to an $n$-immanon-state is non-trivial, since it entails the choice of symmetry that the $n+1$-immanon-state shall fulfil. The dynamics of immanons in exemplary situations -- e.g.~in a harmonic potential or on a lattice -- could then be studied and compared to the well-known behavior of bosons and fermions. In other words, even the most basic many-immanon-theory remains to be formulated coherently and rigorously. Such theory would allow to tackle speculative questions: Could the hitherto seemingly perfect pairwise exchange symmetry be broken above a certain threshold, and non-pairwise immanonic symmetry  already be realized in Nature? Could certain allegedly fermionic or bosonic particles be near-fermionic or near-bosonic immanons?

This seems to us a worthwhile future agenda, since, complementing the theoretical motivations for their study, the implementation of few-immanon interference experiments is in reach with current technology: In order to implement three immanons in the laboratory, a three-dimensional auxiliary Hilbert space is necessary to implement the three-wise exchange symmetry. While polarization as a two-dimensional degree of freedom is not sufficient, the combination with other degrees of freedom under good control, such as, e.g., the time-of-arrival, immediately gives access to the necessary three-dimensional internal Hilbert space. Photonic experiments confirming, e.g.~the partial Pauli principle for the $\lambda=(2,1)$-immanons and the seed-dependence reflected by Eq.~(\ref{notinvariantP}), are therefore feasible.

The relationship between immanons -- which are discrete intermediaries between bosons and fermions -- with quons \cite{PhysRevD.43.4111} -- which lie on the continuous transition between the two extremal species -- is another immediate theoretical desideratum. It also remains to be explored how many-immanon-states collectively interfere in general, along the lines of \cite{Tichy:2012NJP}, beyond bunching phenomena discussed in Section \ref{bunching}. Next to the case of pure states, the question arises how immanons behave at finite temperatures, e.g.~to which extent a many-immanon-gas can condense.

Studying immanons may also lead to new insights in other fields: The characterization of entangled states is  alleviated considerably when restricted to  states that obey certain symmetries. The sector of fully permutation-symmetric \cite{Aulbach2010,Markham2011,Hubener2009a,Martin2010,Baguette2014} and fully permutation anti-symmetric states \cite{PhysRevA.77.012104,Christandl:2012aa} are well-known and understood, as well as the differences between bosons and fermions with respect to their potential to generate entangled states by propagation and detection \cite{tichy:2013PRA}. The current results motivate the characterization of entangled states that fulfil immanonic symmetries.

The very simulation of the scattering of many bosons is a hard compuational problem, known as boson-sampling. The fermionic counterpart, involving fast-to-compute determinants instead of hard permanents, is efficiently solvable. We expect that the ladder of increasing complexity for immanants \cite{buergisser} is inherited by the respective sampling problem, i.e.~immanon-sampling will be of intermediate complexity. Generalizing Ref.~\cite{Troyansky:1996ve} to immanons, the immanant of a general matrix can be written as the quantum expectation value of an experiment -- we expect that the large variance on the expectation value that impedes the efficient measurement of the permanent as a quantum expectation value will also jeopardize the quantum computation of immanants.

In general, the current interplay  \cite{1751-8121-49-9-09LT01} between the physics of many-particle interference \cite{PhysRevA.91.013811,Tichy:2010ZT,Tichy:2012NJP}, the mathematics of immanants \cite{Brylinksi} and the computational complexity gap between determinants and permanents \cite{Aaronson:2011kx} promises further cross-fertilisation and interesting interdisciplinary insights.

\subsection*{Acknowledgements} The authors acknowledge financial support  from the Villum Foundation. They thank Hubert de Guise for pointing out a mistake in an earlier version of the manuscript and for insightful comments and discussions, and Suvrit Sra for making them aware of the Schur orthogonality relation,  Eq.~(\ref{schurortho}). 

%\bibliographystyle{h-physrev}
%\bibliography{LitPhysPerm}

\end{document}